\documentclass[%
 reprint,
 amsmath,amssymb,
 aps,
]{revtex4-1}
\usepackage{graphicx}
\usepackage{dcolumn}
\usepackage{bm}
\usepackage{verbatim}
\usepackage{epsfig}
\usepackage{amsmath,color}
\usepackage{amssymb}
\usepackage{psfrag}
\usepackage[T1]{fontenc}
\usepackage{graphicx}
\usepackage[usenames,dvipsnames]{pstricks}
\usepackage{pst-grad} 
\usepackage{pst-plot} 
\usepackage{algorithmic}

\usepackage{afterpage}

\usepackage{amsthm}

\theoremstyle{remark}

\theoremstyle{definition}

\newcommand{\hslashslash}{%
  \raisebox{.9ex}{%
    \scalebox{.7}{%
      \rotatebox[origin=c]{18}{$-$}%
    }%
  }%
}

\newcommand{\dslash}{%
  {%
   \vphantom{d}%
   \ooalign{\kern.05em\smash{\hslashslash}\hidewidth\cr$d$\cr}%
   \kern.05em
  }%
}

\begin{document}

\preprint{APS/123-QED}

\title{Hindering loads prompt clustered configurations that enhance stability during cargo transport by multiple Kinesin-1}



\author{Shreyas Bhaban, Rachit Srivastava, James Melbourne, Saurav Talukdar, and Murti V. Salapaka}
\affiliation{
University of Minnesota, Minneapolis, USA.
}%





\begin{abstract}
Transport of intracellular cargo is often mediated by teams of molecular motors that function in a chaotic environment under varying conditions. We 
show that the motors have unique steady state behavior which enables transport modalities that are robust. Under reduced ATP concentrations, multi-motor configurations are preferred over single motors. Higher load force drives motors to cluster, but very high loads compel them to separate in a manner that promotes immediate cargo movement once the load subsides. These inferences, backed by analytical guarantees, provide unique insights into the coordination strategies adopted by molecular motors to transport intracellular cargo. 
\end{abstract}

\maketitle

\section{Introduction:}

Intracellular cargo such as vesicles, filaments and organelles are transported inside the cell by molecular motors like kinesin, dynein and myosin. The motors convert chemical energy to mechanical energy through ATP hydrolysis \cite{schnitzer1997kinesin} and traverse on microtubule filaments, while operating individually as well as in homologous or heterologous teams \cite{kural2005kinesin,leduc2007detection}. Numerous experimental studies have analyzed multi-motor ensembles \cite{jamison2012cooperative,rai2022multimodal} and have demonstrated the benefits of working in teams, such as enhanced run-lengths \cite{kunwar2008stepping,muller2010bidirectional} and robustness in a turbulent environment \cite{kunwar2010robust}. On the analytical side, there are multiple studies that investigate cargo transport by two \cite{berger2012distinct,driver2010coupling} or more than two \cite{kunwar2010robust} motors and analyze average velocity, run-length and distributions of bound motors \cite{klumpp2005cooperative} that shed light on team behavior. Several of these studies utilize a probabilistic description of single motor behavior obtained experimentally \cite{mehta1999single, niekamp2021three, shrivastava2019stiffness,shrivastava2022cargo}. They often rely on data obtained from instruments such as optical tweezers \cite{visscher199838, mallik2009intracellular,bhaban2018single, roychowdhury2013design} to build and analyze models that describe transport of cargo by a team \cite{klumpp2015molecular,klumpp2005cooperative,posta2009enhancement,kunwar2010robust,bhaban2016interrogating,talukdar2016steady, shrivastava2018transport,shrivastava2019semi}. These models, coupled with Monte-Carlo (MC) methods, offer numerous benefits in terms of simulating behavior of teams of motors that help guide experiments that are insightful but demanding in terms of time and resources \cite{kunwar2008stepping}. While MC simulations have proven helpful, they suffer from the following drawbacks - (i) The MC simulations are unable to detect modes/methodologies of transport that are rare. (ii) If a specific mode of transport is observed in a realization of MC simulation, it is difficult to further deduce the details of that mode. (iii) The MC simulations are unable to provide insights into the asymptotic behavior of the team of motors. 
On the other hand, simplified models that encapsulate the dynamics of team of motors provide insights analytically. However, these models lie on the other end of the spectrum to MC simulations where a detailed description is not addressed and simplifying assumptions may take the simulations farther away from reality. 

In this article we adopt a semi-analytic methodology \cite{materassi2013exact} that, while capturing the modeling depth afforded by the MC methods, is able to provide conclusions on asymptotic behavior and a deductive capability that is lacking in MC methods. Here, the method utilizes a projection of an underlying infinite dimentional Markov model to produce a finite dimensional Markov model. The resulting model enables an \textit{exact} calculation of the probability distribution function of the behavior of motors relative to each other while transporting cargo in a team, through a computationally efficient semi-analytic approach.
We establish (for a generalized case of a team of finite motors) that the relative configurations of the motors in the team while transporting cargo, have a \textit{unique and non-trivial} steady state distribution. The uniqueness of the distribution points to the robustness of the system of multiple motors carrying a cargo, where irrespective of the initial arrangement, the motors in the team assume a unique orientation. Moreover the distribution, which is dependent on external environmental factors, is determinable; thereby providing a means to reach conclusions on how the team of motors overcomes an adverse environment. Here the efficacy of the approach is demonstrated by investigating how teams deal with changing ATP concentrations and external load forces by examining two and three motor ensembles. Key analytical results reported here are that, as ATP concentrations are lowered, teams surprisingly tend to favour multi-motor configurations over single motor configurations; with the cargo more probable to be carried by more than one motor than only a  single motor. The implications are that as the ATP concentration reduces, the average run-length of multi-motor ensembles increases even though average ensemble velocity decreases. 
It is known that reduced ATP concentration lead to reduced single motor velocity, which is hypothesized as a reason for increased runlengths  when cargo is transported by two motors \cite{xu2012tuning}. Such a mechanism also provides a possible regulating mechanism by which transport occurs by an ensemble containing a mixture of different species of motors \cite{pan2006mechanism}.

The analysis presented in this article also provides an explanation for the emergence of various multi-motor configurations (based on the properties of the individual motors in the ensemble), as a consequence of changing load forces on the cargo. When the cargo is subjected to increasing load forces, motors adopt a form of cooperation by clustering together in order to handle higher load forces. The propensity to cluster increases with load force. However the trend (of closer clustering to handle more load forces) does not hold for all loads. Indeed, for very high values of load forces (possible when cargo encounters obstacles along the path of travel), the teams abandon clustering. They instead resort to spreading out in a manner where the entirety of the load falls on as few motors as possible, with the rest of the motors assuming configurations where they are subjected to forces that are near the 'stall-force' of a single motor (which is the maximum force against which a motor can actvely step on the microtubule). 
We explain the observations by analyzing how the increasing load force is shared as the motors orient themselves to cluster/spread out. The motors clustering together to handle increasing load forces facilitates a more equal sharing of load force on the cargo. We propose that it prevents the motor from being loaded beyond stall force, thereby maintaining them under stalling condition and aiding forward cargo motion. On the contrary, preference for non-clustered configurations might lead to a more unequal load sharing, thereby increasing the likelihood of having the farthest stretched motor/s being loaded beyond their stall force. In the case that the load force is very high where even with equitable load sharing the force on each motor is beyond its stall force, the motor ensemble abandons clustering. It prefers configurations where most of the load is taken by the least possible number of motors, while other motors are only loaded to be near stall and are primed to take steps after a small reduction of load force. 
Such a separation of motors at very high loads is possibly preferred in order to enable higher probability of immediate cargo translation in the forward direction, once the high forces have elapsed and the high load phenomenon has subsided.

In this article, we begin with a brief overview of the semi-analytic model and introduce terminology to establish the finite-dimensional model. We then establish the existence and uniqueness of the steady state distribution of various configurations of the motors in a team, by instantiating the model for molecular motors involving stepping, detachment and reattachment probabilities. We subsequently analyze the impact of changing ATP concentrations and load forces on ensembles of multiple  Kinesin-I motors by studying their effect on the steady state distribution. We leverage the knowledge of the unique steady state distribution to explain the ovserbed behavior of motor ensembles during cargo transport. 




\section{Semi-analytic Model:}

\begin{figure}
	\centering
	\includegraphics[scale = 0.35]{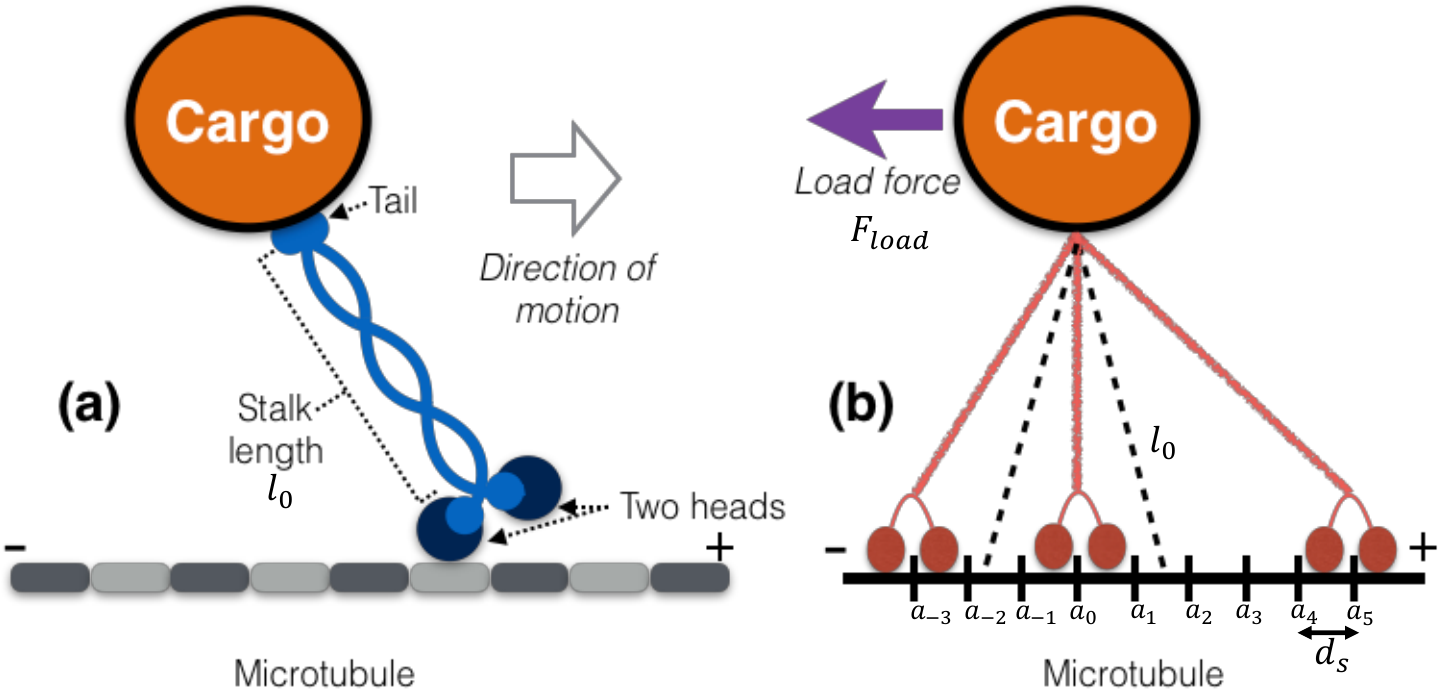}
	\caption{(a) Schematic of structure of the molecular motor Kinesin-I (b) Representation of an ensemble of three motors carrying a cargo. $F_{load}$ is the load force opposing the cargo. 
	The relative configuration is $\sigma_t = (-3, 0, 5)$ which is symbolically represented as $M|||M|||||M$, where $M$ denotes motor location on the microtubule and | denotes the microtubule locations.}
	\label{fig:multiple_motors}
\end{figure}

For a single molecular motor, the linkage between the motor-head and tail is assumed to behave like a hookean spring when stretched, which offers no resistance when compressed \cite{kunwar2008stepping}. It has a stiffness constant of $K_e (pN/nm)$ and rest length of $l_0 (nm)$ (e.g. see schematic for Kinesin-I motor in Fig. \ref{fig:multiple_motors}(a)). The motor exerts a force $F$ on the cargo which is related to the the linkage length $l$ as,
\[ F(l) = \left\{ 
\begin{array}{l l}
K_{e}(l-l_0) & \quad \text{if} \: l \geq l_0,\\
0 & \quad \text{if} \: |l| < l_0,\\
K_{e}(l+l_0) & \quad \text{if} \: l \leq -l_0.\\ 
\end{array} \right.\]
 
Here, $l \geq l_0$ denotes linkage extension in the direction of the motor motion and ahead of the cargo position while $l \leq -l_0$ denotes denotes linkage extension opposite to the direction of the motor motion and behind of the cargo position. If the motor is subjected to a force, due to the stretching of its linkage, that is higher than the stalling force $F_S$ for the motor then the motor becomes stalled and is unable to take a forward step. The stalling force is an experimentally measurable property of the motor protein and has been measured using probing instruments such as optical tweezers \cite{svoboda1994force}. We consider the case where the cargo is subjected to a constant load force $F_{load}$ which opposes the motion of the motors. We define the mean position of the cargo $X_C$ as the equilibrium position obtained when the forces exerted by the motors through their linkages and the load force $F_{load}$ balance each other. The cargo position is assumed to have a standard deviation of $\sigma_{th}$ which captures the thermal fluctuations of the cargo during transport. When there are no motors engaged to the microtubule, the cargo is considered to be "lost".

The microtubule filament over which the motor traverses, is modeled as a sequence of equally spaced dimers $a_q = a_0+qd_s$, where $a_q$ is the location of $q^{th}$ dimer and $d_s$ is the dimer length. While defining the model for multiple motors carrying a common cargo on the microtubule filament, we make the following assumptions:
\begin{itemize}
	\item The cargo is carried by an ensemble of $\bar{m}$ motors
	\item The motors $\bar{m}$ are irreversibly bound to the cargo
	\item The cargo is subjected to a constant load force $F_{load}$
	\item Motors attached to the cargo, that are not attached to the microtubule filament, can only reattach to locations on the microtubule that are within the length $l_0$ from the cargo position.	
	\item All the motors are attached to the cargo at the same location, and multiple motors are able to share the same location on the microtubule
\end{itemize}

The locations of motors on the microtubule are represented by a sequence of natural numbers $Z := \{ z_q \}, q \in I$ where $z_q$ is the number of motors are located on the $a_q^{th}$ location of the microtubule and $I$ is the set of integers \cite{materassi2013exact}. Since the typical lengths of microtubule filement (several mircometers)  is several orders of magnitude greater than the average distance traveled by motor proteins (nanometers), $Z$ is effectively a bi-infinite sequence of numbers. Thus, $Z$ represents the \textit{absolute configuration} of the motors on the microtubule. For example, in Fig. \ref{fig:multiple_motors}(b) the absolute configuration is $Z=[...~ 1~ 0~ 0~ 1~ 0~ 0~ 0~ 0~ 1~ ...]$. Each of the motors can take a forward step to the next location, detach from its current location or can reattach to a new location on the microtubule. When a stepping, detachment or reattachment event occurs, the configuration changes from $Z$ to $Z'$; with the transition rate between $Z$ and $Z'$ being denoted by $\lambda_Z(Z',Z)$ (see Appendix for details on obtaining the rates). For e.g. if the motor at the location $a_0$ in Fig. \ref{fig:multiple_motors}(b) takes a forward step, the new configuration would be $Z=[...~ 1~ 0~ 0~ 0~ 1~ 0~ 0~ 0~ 1~ ...]$. In this case, $\lambda_Z(Z',Z)$ would represent the probability rate of stepping of the motor at location $a_0$. 

With the knowledge of the transition rates, it is possible to define an infinite dimensional Markov Model similar to the one described in \cite{doob1945markoff, gillespie1977exact}.  Here, the probability of going from $Z$ to $Z'$ in time $\Delta t$ is represented by  $\lambda_Z(Z',Z) \Delta t$. Thereby, the probability that the configuration is $Z'$ at $t$ given that it was $Z$ at initial time $t_0$, is represented by $P_Z(Z, t|\bar{Z}, t_0)$, and satisfies the probability Master Equation where the rates are given by $\lambda_Z$. For small time intervals $\Delta t$, note that $P_Z(Z', t + \Delta t|Z, t) = \lambda_Z(Z', Z)\Delta t$. However, as mentioned previously, the infinite dimentionality of the Master Equation based on absolute configurations $Z$ makes it intractable.  

The issue is resolved in \cite{materassi2013exact} by showing that the maximum distance between the forward most (vanguard) and rear most (rearguard) motor is bounded. When $\bar{m}$ motors carry a common cargo subjected to a load force $F_{load}$, the maximum distance is shown to be bounded by $S_{max} = max  \left\{ \frac{\bar{m}F_s-F_{load}}{K_{e}} +d_s, \frac{F_{load}}{K_e} \right \} + 2l_0$, where $d_s$ is the step size for the motor (e.g. $d_s=8~nm$ for Kinesin-I ). This enables the projection of a bi-infinite absolute configuration $Z$ to a finite dimensional \textit{relative configuration} denoted by $\sigma$. The relative configuration represents the \textit{relative} locations of all the motors in an ensmble carrying the common cargo. For a symbolic representation, $M$ denotes a motor attached to a location on the microtubule and $|$ denotes distinct microtubule locations.


For example, the ensemble in Fig. \ref{fig:multiple_motors}(b) can be represented by the relative configuration $M|||M|||||M$. To enhance the symbolic redability of relative configurations with large extents, we represent such relatve configurations with "-(the number of locations between motors)-" instead of equivalent number of $|'s$. For example, $M||||||||||M$ is represented by $M-(10)-M$.

Though, in general, projections do not preserve the Markov property of a model, it is shown in \cite{materassi2013exact} that upon projecting absolute configurations onto relative configurations, the resulting model retains the Markov property. Thereby, the transition rates between relative configurations $\sigma$ and $\sigma'$, denoted by $\lambda_\sigma(\sigma',\sigma)$, can be computed using the transition rates between the corresponding absolute configurations $\lambda_Z(Z',Z)$ (\cite{materassi2013exact}, equation 5).  Thus, the probability of the system being in the relative configuration $\sigma$ at time $t$, given by $P_\sigma(\sigma,t)$, obeys the Master Equation :

\begin{align*}
\frac{\partial}{\partial t}P_{\sigma}(\sigma,t) &=\displaystyle\sum_{\sigma' \in~ \bar{S}} \lambda_{\sigma}(\sigma,\sigma')P_{\sigma}(\sigma',t)\\ &-P_{\sigma}(\sigma,t)\displaystyle\sum_{\sigma' \in~ \bar{S}} \lambda_{\sigma}(\sigma',\sigma)
\end{align*}
where $\bar{S}$ is the set of relative configurations. For an ensemble of a finite number of motors $\bar{m}$ carrying a common cargo while subjected to a finite load force $F_L$, the relative configuration space $\bar{S}$ is finite \cite{materassi2013exact}. Let the total number of relative configurations in $\bar{S}$ be $\bar{n}$, thereby $\bar{S}= \{ \sigma_1,\dots, \sigma_{n}, \sigma_{n+1},\dots \sigma_{\tilde{n}} \}$. We can denote the probability vector as $P(t)=[ P_1(t),\dots,P_n(t), P_{n+1}(t),\dots,P_{\tilde{n}}(t)]$ where each element $P_i(t) = P_{\sigma}(\sigma_i,t)$ i.e. the probability of the system being in the $i^{th}$ relative configuration, $\sigma_i$, at time $t$. The probability vector $P(t)$ satisfies the master equation 
\begin{equation}\label{eq:Prob_propogate}
\frac{d}{dt} P(t)=\mathbf{\bar{\Gamma}}P(t)
\end{equation}
where $\mathbf{\bar{\Gamma}} \in \mathbf{R}^{\tilde{n} \times \tilde{n}}$ is defined by the transition rates $\lambda_{\sigma}(\sigma_j,\sigma_i)$.   The transition rates are obtained from the transition rates between corresponding absolute configurations $\lambda_Z (Z_j,Z_i)$ that correspond to stepping, detachment or reattachment events and are influenced by external conditions such as $F_{load}$ and ATP concentration. Solving for $P(t)$, we get $P(t) = e^{\mathbf{\bar{\Gamma}}(t-t_0)}P(t_0)
$,
where $P(t_0)$ is an initial probability vector. 

In order to obtain a model amenable to simulation, we consider a standard discrete time approximation of a continuous Markov process \cite{gardiner1985handbook}. 
For a fixed time interval $\Delta t$, we define a time homogeneous Markov chain by the transition probabilities
\begin{align} \label{eq: discrete transition matrix}
    P(\sigma',\sigma) &= \begin{cases}
                    \Delta t \ \lambda(\sigma', \sigma) &\hbox{ for } \sigma' \neq \sigma
        \\
        1 - \Delta t \sum_{\sigma' \neq \sigma} \lambda(\sigma', \sigma) &\hbox{ for } \sigma' = \sigma.
    \end{cases}
\end{align}
and the transition matrix
\begin{align} \label{eq: transitionMatrixJ}
\mathbf{J} = (P(\sigma',\sigma))_{\sigma,\sigma' \in \bar{S}}
\end{align}
The time interval $\Delta t$ is chosen to be small enough that only a single event, i.e. single motor stepping, detachment or reattachment, can occur. Therefore the probability of simultaneous occurrence of more than one event is negligible.  In this case the continuous process $(\sigma, t)$ and discrete approximation $\sigma_n := (\sigma, n)$ defined through the time homogeneous transition probabilities in \eqref{eq: discrete transition matrix}  are in close agreement as $P_\sigma(\sigma', (n+1) \Delta t|\sigma, n \Delta t) \approx  \lambda_\sigma (\sigma', \sigma) \Delta t =: P_\sigma(\sigma'_{n+1}, \sigma_n)$. 

In the following section we will condition this discrete time Markov chain, on the event that a fixed minimal number of motors remain attached to the microtubule. This would enable us use the numerical simulations, to derive biological insights into the collaborative behavior of teams of motors under varying external conditions.

\section{Unique Steady State Distribution: } 


When a cargo being carried by the ensemble of $\bar{m}$ motors traverses along the microtubule, eventually all the motors would permanently disengage from the microtubule and the cargo will be lost. Indeed, if we propagate the probability vector $P(n)$ using equation (\ref{eq: discrete transition matrix}), the steady state distribution $P_{ss}$ indicates that cargo being lost is the eventual outcome. Thus, if $\phi$ denotes the relative configuration where no motors are attached to the microtubule (which characterizes the state where the cargo is permanently lost) then the nature of $P_{ss}$ is such that 
\begin{align} \label{eq: Pss_nature}
P_{ss} = \begin{cases}
1 &\hbox{ if } \sigma=\phi
\\
0 &\hbox{ otherwise }.
\end{cases}
\end{align}
Although this corresponds to an intuitive fact since loss of cargo is an inevitable eventual outcome of transport of cargo \cite{kunwar2008stepping}, not much information regarding the relative arrangements of the motors can be gathered by using such a probability distribution. We therefore introduce a non-trivial notion of a `steady state', which corresponds to a conditional probability distribution where the cargo remains attached to the microtubule through a finite number of motors. Under such a conditioning, it is not evident apriori what the nature of $P(n)$ would be like; whether it has a unique steady state or not, whether the steady state is unique or not  and whether the steady state captures the dynamics of the original state space $\bar{S}$. By using the underlying model for molecular motors \cite{kunwar2008stepping} and a constant load force on the cargo, we utilized properties of the underlying Markov model to prove the following :

\textit{Consider a Markov Model with a state space $S$, transition matrix $\mathbf{J}$ \eqref{eq: transitionMatrixJ} for an ensemble of $\bar{m}$ motors carrying a common cargo; conditioned on the event that at least $m \in [1,\bar{m}-1]$ motors always remaining attached to the microtubule and the cargo subjected to constant load force $F_{load}$. Then, the associated Markov chain has a unique steady state distribution $P_{ss}$.}

The proof is provided in the supplementary material at \cite{bhaban2018SuppMaterial} section I theorems I.1 to I.12. Here, we begin with an ensemble of $\bar{m}$ motors carrying a common cargo, with the state space $S = \{ \sigma_1,\dots, \sigma_{n} \}$. Here, $S$ contains all the relative configurations with at least $m \in [1,\bar{m}-1]$ motors attached to the microtubule. Thereafter -
\begin{itemize}
	\item Using theorems I.1-I.4, we use the properties of single motor and multiple motor ensembles to first show that irrespective of the initial arrangement of the ensemble of finite $\bar{m}$ motors, it is possible to reach a single relative configuration through finite events of motor stepping, detachment or reattachment. 
	\item Consequently, in theorems I.5-I.12, we utilize the aforementioned single relative configuration to build the irreducible and aperiodic state space $S$ that has a unique steady state distribution $P_{ss}$. 
\end{itemize}

It would be of interest to investigate the conditions where a known number of motors in the ensemble remain attached to the microtubule at a given time instant i.e. $m = 0, 1, 2, 3, \dots$; but we do not pursue it here. However, it is reasonably straightforward to distinguish the condition of the cargo not being lost ($m \geq 1$)  from cargo being lost ($m=0$). For the remainder of this article, we analyze results pertaining to the case where the cargo that is being carried by an ensemble of $\bar{m}$ motors is not lost. This means that the cargo remains attached to the microtubule through at least one motor i.e. $m \geq 1$. Based on the proof provided in supplementary material \cite{bhaban2018SuppMaterial} section I, we know that a unique steady state exists for $m=1$. In \cite{bhaban2018SuppMaterial} section II it is seen that the value of the probability vector $P(n)$ obtained by propagating $P(n+1) = \mathbf{J}P(n)$ with time, approaches $P_{ss}$ obtained by solving $P_{ss}= \mathbf{J}P_{ss}$. Thus by propagating $P(n) = \mathbf{J}^n P(0)$, we can conveniently obtain an estimate for $P_{ss}$ and also analyze the dynamics of the process of orientation of the motors in the ensemble as time progresses, thus making the long term behavior of the ensemble tractable. Furthermore, in \cite{bhaban2018SuppMaterial} section III we use the model to compute important biological quantities governing intracellular traffic, such as average cargo velocity and run-length. Here, we show that an instantiation for a case of two and three-motor ensembles of Kinesin-I yields quantitative and qualitative outcomes that are in good agreement with existing studies. The concurrence between results obtained using the analytic model and existing literature justifies the usage of the semi-analytic framework to arrive at conclusions about the steady state dynamics of the ensemble and how it responds to changing external conditions. 

The existence and uniqueness of the steady state distribution indicates a degree of robustness in the multi-motor cargo transport system. 
Furthermore, the conditional steady state distribution $P_{ss}$ is independent of the initial distribution $P(0)$ at the time $t_0=0$, indicating that no matter how the motors are oriented prior to the initiation of the cargo transport, they prefer to align themselves according to a fixed distribution that is dependent on external conditions. 

In the next sections, we utilize the semi-analytic model to quantify the effect of external conditions on the behavior of the multiple motors, which we observe is captured by the steady state $P_{ss}$. In particular, we analyze the effect of varying ATP concentration and load forces on transport of cargo by multi-motor ensembles.\\

\section{Effect of ATP concentration:}

\begin{figure*}
\centering
\includegraphics[scale = 1.2]{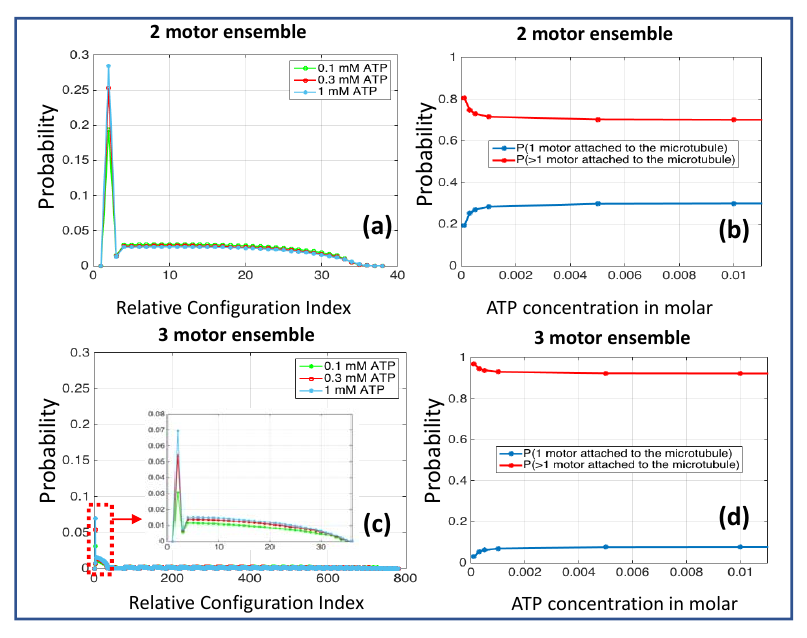}
\caption{(a) Variation of probability distribution function with ATP concentration ($F_{load} = 0.2 pN$) for a 2-motor ensemble. The finite state space is defined as $S = \{ \sigma_i\}, i \in [1,n] = \{ \sigma_1,\dots, \sigma_{n} \}$. Here, we denote $i$ as the 'Relative Configuration Index' which is represented on the x-axis. (b) represents the probability of 1-motor attached and more than 1-motor attached, corresponding to the ATP variation investigated in (a). (c) denotes the variation of probability distribution function with ATP concentration ($F_{load} = 0.2 pN$) for a 3-motor ensemble. (d) represents the probability of 1-motor attached and more than 1-motor attached, corresponding to the ATP variation investigated in (c).} \label{fig:ATPchange_2and3motor}
\end{figure*}

We begin by analyzing the impact of changing ATP concentrations on the steady state probability distribution function (pdf) of the relative configurations for two and three-motor ensembles, at a constant load force. Note that in the relative state space model (section III), we have enforced the condition that the last motor remains attached to the microtubule. We define the finite state space as $S = \{ \sigma_i\}, i \in [1,n] ~ \text{i.e.} ~ S = \{ \sigma_1,\dots, \sigma_{n} \}$. Here, $i$ is the 'Relative Configuration Index'. Note that - in all the pdfs shown in Fig.\ref{fig:ATPchange_2and3motor} - \ref{fig:LoadForce_3motor_High}, we have included the relative configuration $\phi$ (which represents the zero motor attached condition with relative configuration index 1 i.e. $\sigma_1$) for the sake of completeness. However, since we analyze the conditional pdfs with the condition that a single motor remain attached to the microtubule, the probability $P_{ss}(\sigma_1)=P_{ss}(\phi)=0$. 

Fig. \ref{fig:ATPchange_2and3motor} (a-d) shows the impact of ATP concentration on two and three motor ensembles. For the two and three motor configurations, we observe that the pdf is impacted by changing ATP concentrations. An observation of Fig. \ref{fig:ATPchange_2and3motor} (a) and (b) would indicate that, the probability of the relative configuration $\sigma_2 = M$ reduces with reducing ATP concentration. To further investigate this, we analyzed the probability of $1$ motor being attached to the microtubule (denoted by $P(\sigma_2)$) and the probability of more than $1$ motor being attached to the microtubule (equal to $1-P(\sigma_2)$ ) over multiple ATP concentrations, as shown in Fig. \ref{fig:ATPchange_2and3motor}  (b) and (d). For both two and three motor ensembles, with reducing ATP concentration, we observe a reduction in their respective $P(\sigma_2)$ and an increase in $1-P(\sigma_2)$. Therefore, it can be said that with reducing ATP concentration, motors are less likely to adapt a single motor configuration (i.e. $\sigma_2=M$) and more likely to adapt multi motor configurations.

Let us examine this by tracing the impact of ATP concentration on single motor transition rates, based on the single motor model for Kinesin-I utilized in this article (see appendix). We observe that reduction in the ATP concentration reduces the stepping rate, which in turn impacts the detachment rate of the single motor. The motor reattachment rate is assumed to be independent of the ATP concentration. Thus, reducing ATP concentration would reduce the motor detachment rates, but leave the reattachment rates unchanged. The combined impact of this is that configurations with more than one motors attached to the microtubule are less likely to undergo detachment events and more lilely to remain in configurations where multiple motors remain attached to the microtubule. 
Thus, as evidenced in Fig. \ref{fig:ATPchange_2and3motor}  (b) and (d), reduction in ATP concentration leads to an increase in the probaility of multiple motors remaining attached to the microtubule. 

\paragraph*{Impact}: In a multi-motor ensemble, a higher probability of multi-motor configurations would indicate an increased contribution of the other motors in the ensemble. This would lead to an increased participation of $2^{nd}$ motor, $3^{rd}$ motor, and so on in the overall transport of the cargo. The impact of increased participation of other motors on overall transport can be seen in (\cite{bhaban2018SuppMaterial} section IV) where, with reducing ATP concentrations, the average velocity for multi-motor ensembles \textit{reduces} but the runlength \textit{increases}. The increase in runlength can be attributed to the fact that, even though the velocity has diminished with reducing ATP concentration, the probability of more than one motor remaining attached to the microtubule has increased. Here, the cargo has a higher probability of remaining engaged to the microtubule. This enables the second motor (and third motor for three motor ensemble) to contribute to the cargo motion, allowing for a higher probability of the cargo being linked to the microtubule during the course of cargo travel. It enhances the overall distance covered by the multi-motor ensemble and corroborates experimental observations such as \cite{xu2012tuning}, while providing an explanation based on steady state probability distributions of the multi-motor ensembles. Our analysis predicts that for multi-motor ensembles, the probability of cargo remaining attached through more than one motor increases with reduced ATP concentrations contributing to increased cargo run-length.



\section{Effect of load force}

\begin{figure*}
	\centering
	\includegraphics[scale = 0.65]{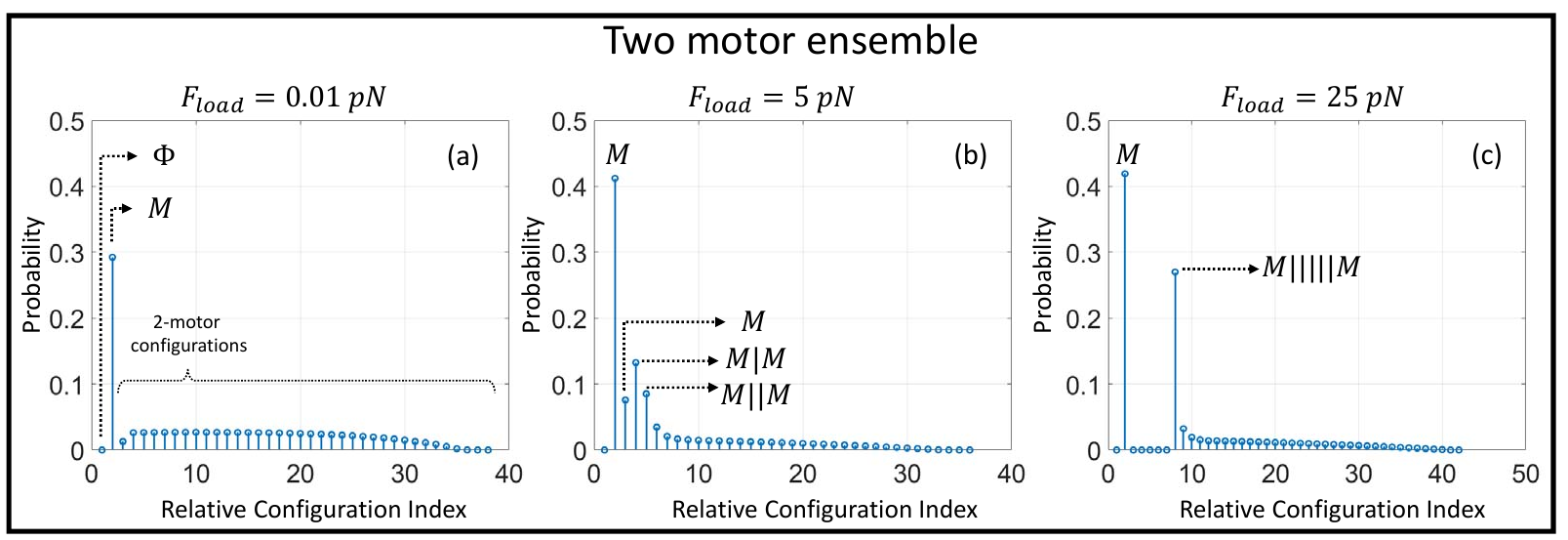}
	\caption{Variation of steady state probability distribution $P_{ss}$ for a two-motor ensemble with load force (a) $F_{load} = 0.01~pN$ (b) $F_{load} = 5~pN$ (c) $F_{load} = 25~pN$ . For a relative configuration $\sigma_i$, we denote $'i'$ as the 'Relative Configuration Index'. The relative configurations $\sigma_1=\phi$, $\sigma_2 = M$, and $\sigma_3, \sigma_4, \dots$ all represent two-motor configurations.} \label{fig:LoadForce_2motor}
\end{figure*}

\begin{figure*}
	\centering
	\includegraphics[scale = 0.6]{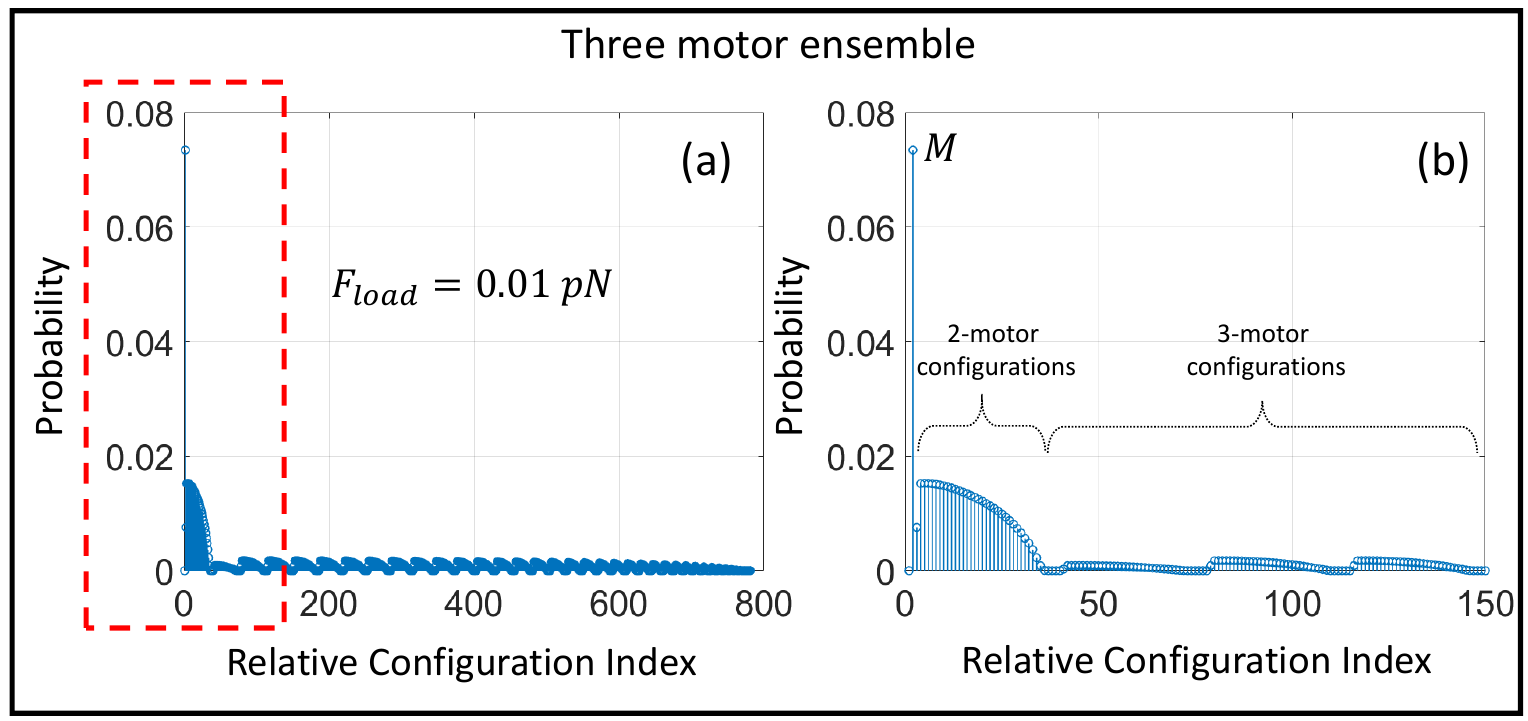}
	\caption{(a) Steady state probability distribution $P_{ss}$ for a three-motor ensemble at low load force $F_{load} = 0.01 pN$. For a relative configuration $\sigma_i$, we denote $'i'$ as the 'Relative Configuration Index'. The relative configurations $\sigma_1=\phi$, $\sigma_2 = M$, $\sigma_3-\sigma_{41}$ represent two-motor configurations, and $\sigma_{42}, \sigma_{43}, \dots$ all represent three-motor configurations. (b) shows the zoomed in plot for $i=1$ to $150$.} \label{fig:LoadForce_3motor_Low}
\end{figure*}

\begin{figure*}
	\centering
	\includegraphics[scale = 0.5]{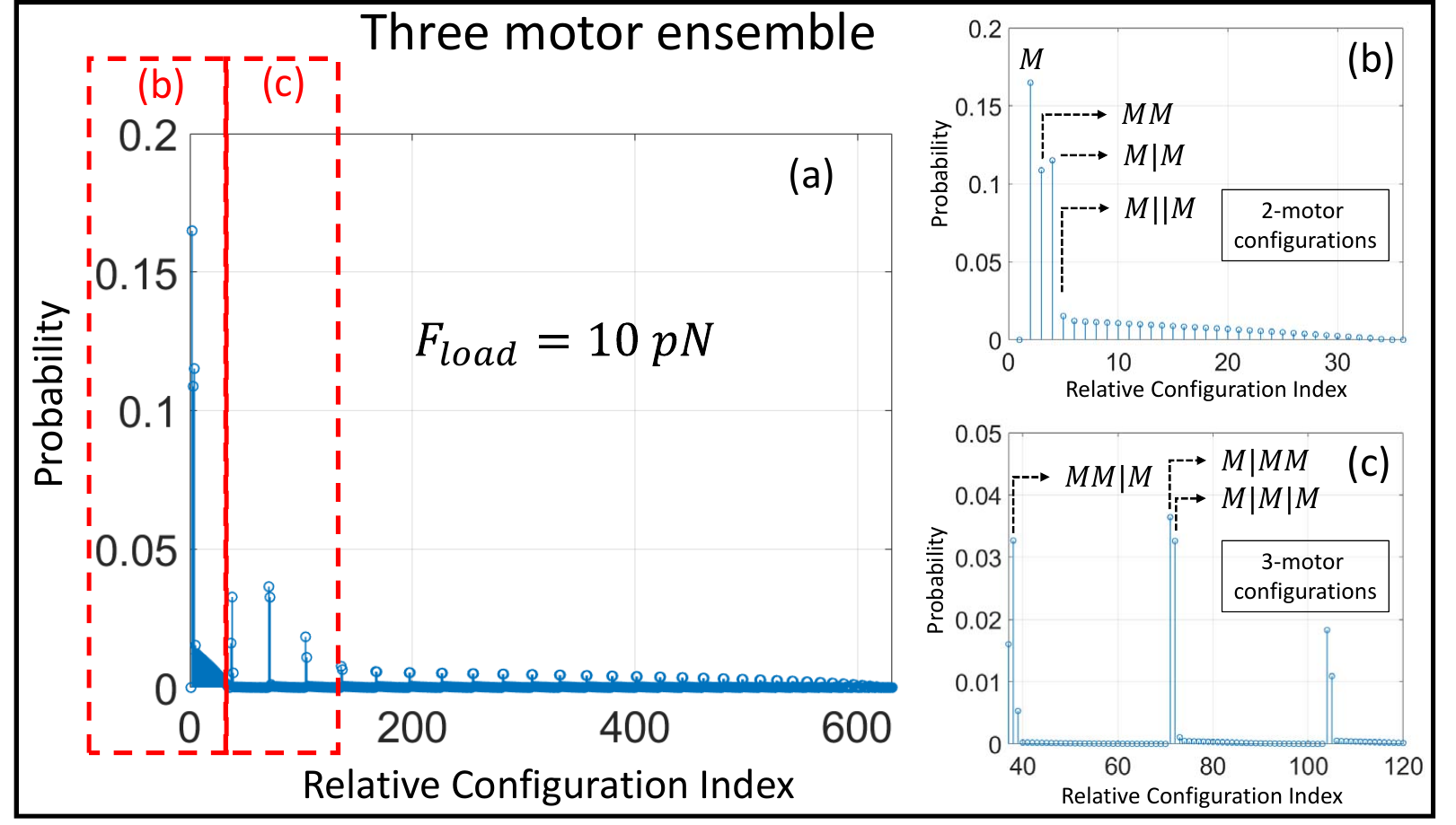}
	\caption{(a)Steady state probability distribution $P_{ss}$ for a three-motor ensemble at mid load force $F_{load} = 10 pN$. For a relative configuration $\sigma_i$, we denote $'i'$ as the 'Relative Configuration Index'. The relative configurations $\sigma_1=\phi$, $\sigma_2 = M$, $\sigma_3-\sigma_{36}$ represent two-motor configurations, and $\sigma_{37}, \sigma_{38}, \dots$ all represent three-motor configurations.(b) shows the zoomed in plot for $i=1$ to $36$ and i.e. 1 and 2-motor configurations (c) shows the zoomed in plot for $i=7$ to $120$ i.e. 3-motor configurations.} \label{fig:LoadForce_3motor_Mid}
\end{figure*}

\begin{figure*}
	\centering
	\includegraphics[scale = 0.6]{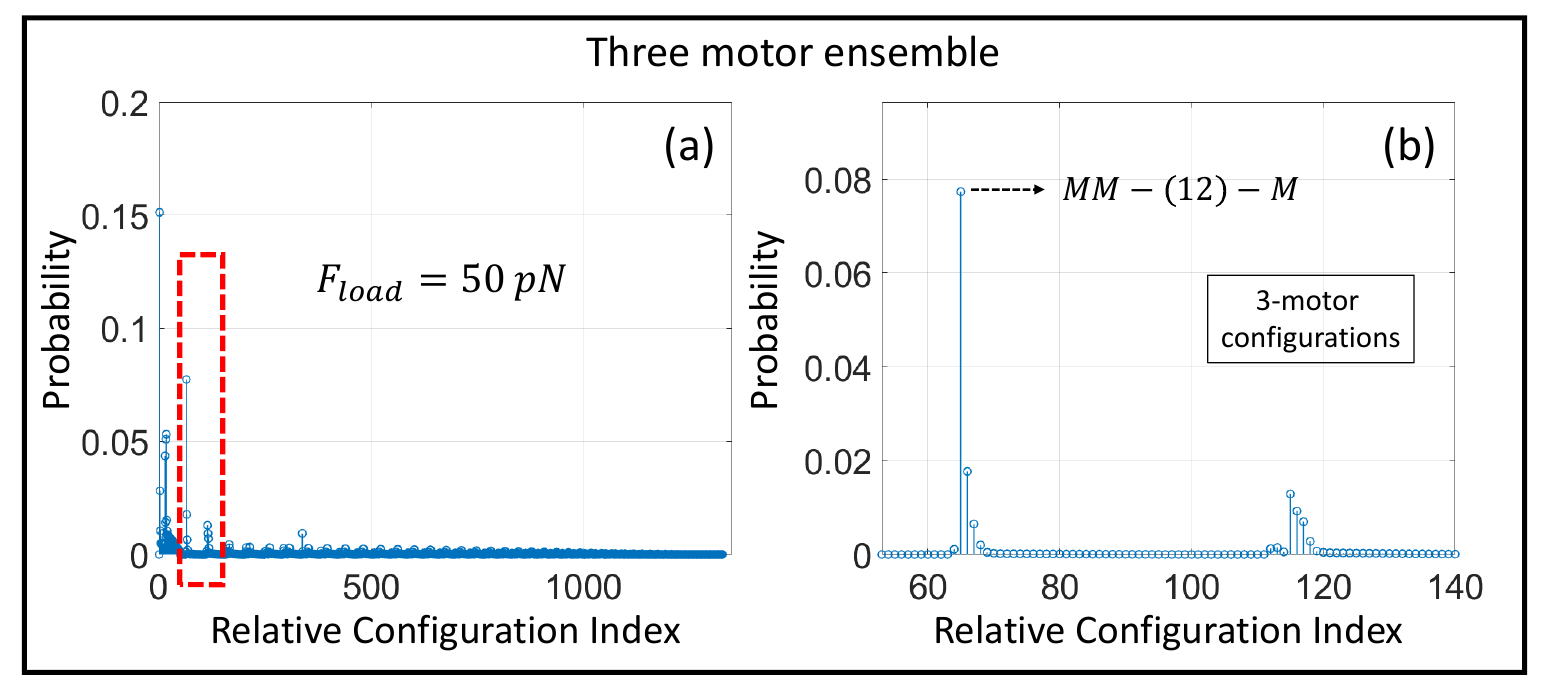}
	\caption{(a) Steady state probability distribution $P_{ss}$ for a three-motor ensemble at high load force $F_{load} = 50 pN$. For a relative configuration $\sigma_i$, we denote $'i'$ as the 'Relative Configuration Index'. The relative configurations $\sigma_1=\phi$, $\sigma_2 = M$, $\sigma_3-\sigma_{52}$ represent two-motor configurations, and $\sigma_{53}, \sigma_{54}, \dots$ all represent three-motor configurations. (b) shows the zoomed in plot for the red-colored section outlined in (a) i.e. $i=53$ to $140$ for the 3-motor configurations.} \label{fig:LoadForce_3motor_High}
\end{figure*}

\begin{figure*}
	\centering
	\includegraphics[scale = 1]{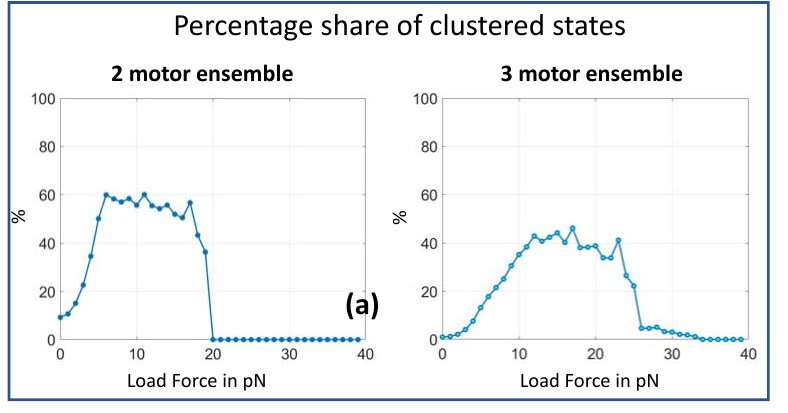}
	\caption{Percent share of probability of 'clustered states' among the probability of remaining states in the steady state distribution $P_{ss}$, with varying load force $F_{load}$ for (a) 2-motor ensembles and (b) 3-motor ensembles.} \label{fig:PerClusteredStates_2and3motor}
\end{figure*}

We further analyze the impact of varying load forces on the steady state probabilities of relative configurations of two and three motor ensembles, as shown in Fig. \ref{fig:LoadForce_2motor}- \ref{fig:LoadForce_3motor_High} . 
It is seen that at very small values of load forces a significant majority of configurations have low probabilities with little variation. For e.g. in Fig. \ref{fig:LoadForce_2motor} (a) the probabilities of two motor configurations show less variation. Similar observations can be made for two and three motor configurations in Fig. \ref{fig:LoadForce_3motor_Low} (a) and (b). No one relative arrangement is particularly favored, with the variation in probabilities being very gradual. This indicates that the motors do not prefer any particular relative configuration at these values of load forces. An intuitive explaination is that at such low loads, since equal or unequal load sharing cause insignificant variations in the load forces balanced by each motor in the ensemble, there is no advantage to adhere to a specific orientation. Thus in these regimes motors tend to spread out more evenly while transporting the common cargo.


However, as the load force is increased, certain configurations start to become more and more probable. Consider a two-motor ensemble under $F_{load}=5 pN$, where the steady state distribution for all the configurations is shown in Fig. \ref{fig:LoadForce_2motor} (b). As compared to \ref{fig:LoadForce_2motor} (a), it appears that the probabilities of certain configurations are noticably greater than the rest of the configurations. In particular, among the two-motor configurations the most probable configuration is $M|M$ (followed closely by $M||M$ and $MM$) and is noticably likelier than the rest of the two-motor configurations. (Together, these three configurations make up about $50 \%$ of the two-motor configurations). For a two-motor ensemble under $F_{load}=5 pN$, the maximum distance between the vanguard and rearguard motor, $S_{max}$, is $34$ microtubule locations. We observe that in the most probable two-motor configuration $M|M$, the distance between vanguard and rearguard motor is one microtubule location. This is significantly lower than the maximum distance $S_{max}$, which is $34$ microtubule locations in this case. 
Similarly, consider a three-motor ensemble under $F_{load}=10 pN$ (Fig. \ref{fig:LoadForce_3motor_Mid}). We observe that among the three-motor configurations shown in Fig. \ref{fig:LoadForce_3motor_Mid}(c), the three most probable configurations are $M|MM, MM|M$ and $M|M|M$ (in the decending order of probabilities). Together, these three configurations account for about $25.35 \%$ of the three-motor configurations. 
Here, the maximum distance between the vanguard and rearguard motor, $S_{max}$, is $34$ microtubule locations. Here as well, the most probable configuration of $M|MM$ has the distance between vanguard and rearguard motor of one microtubule location, which is significantly lower than the maximum possible distance of $34$ microtubule locations. We thus notice a common attribute - as the load force is increased, ensembles prefer to arrange themselves in configurations where the vanguard and rearguard motors are separated by a distance significantly lesser than the maximum separation possible. Note that, among the two-motor configurations of the three-motor ensemble shown in Fig. \ref{fig:LoadForce_3motor_Mid}(b), the three most probable configurations are $M|M, MM$ and $M||M$ (in the decending order of probabilities). Together, these three configurations account for about $55.12 \%$ of the two-motor configurations.



To further analyze this observation, we introduce the notion of \textit{extent} and \textit{clustered states}. In a given configuration, let the term \textit{extent} indicate the distance between the vanguard and rearguard motor. A \textit{clustered state} is defined as a relative configuration where the \textit{extent} is bounded by two microtubule locations. By this definition, the clustered states in a two-motor ensemble are $MM, M|M, M||M$ and in a three-motor ensemble are $MMM, MM|M, MM||M, M|MM, M||MM, M|M|M$. We use this definition to track the behavior of clustered states as the load force is steadily increased from a low to high value. Fig. \ref{fig:PerClusteredStates_2and3motor}(a) shows the variation of the percentage share of two-motor clustered states out of all the possible two-motor configurations, with changing load force. It is seen that among the states with two-motor configurations, a higher load force typically leads to the clustered states claiming a higher share of the probability thereby also making them more probable. 


\subsection{Why is clustering preferred with increasing load force $F_{load}$?}

One way to examine this behavior is to trace the impact of load force on the rates of transition between the various configurations of a two-motor ensemble, based on the basic properties of the single motor model. The single motor model for Kinesin-I utilized in this article assumes a probability of reattachment which is independent of the load force (see appendix). This leaves stepping and detachment as the two rates that are dependent upon load force. Let us consider an example of a two-motor ensemble under $F_{load}=5 pN$, (Fig. \ref{fig:LoadForce_2motor} (b)). At any time instant during the journey of the cargo and ensemble from source to destination, the motor configuration would either be a one-motor configuration ($M$) or one of the $34$ two-motor configurations. From an initial state as a two-motor configuration, there is always a finite probability of one of the two motors detaching from the microtubule, leading to the one-motor configuration $M$. Thus, we can assume an initial state of the ensemble as $M$. Here, the other floating motor can reattach to any location within $l_0$ distance of the cargo and lead to a two-motor configuration. Once in a two-motor configuration, the remaining possibilities are that one of the two motors could detach or take a step. (1) If detachment occurs, the ensemble reverts back to the initial state $M$ after which there is a finite probability of it coming back to a two-motor configuration through reattachment. Irrespective if the magnitude of the load force, after every detachment event there is a finite probability of the ensemble returning to a two-motor configuration due to the load-force independent nature of reattachment. (2) If stepping occurrs, then one of the following is possible - either the vanguard motor takes a step or rearguard motor takes a step. When the vanguard motor takes a step, the extent increases whereas when the rearguard motor takes a step, the extent decreases. In any configuration where the vanguard and rearguard motors are separated by one or more than one location (e.g. $M|M, M||M, \dots, M-(34)-M $) and the cargo is under a finite load force, the extension of the vanguard motor linkage is always greater than that of the rearguard motor linkage. Consequently, vanguard motor is subjected to a higher load force than the rearguard motor. In these configurations, the probability of the rearguard motor taking a step is always higher than that of the vanguard motor taking a step (in fact, this disparity in stepping probability is higher for configurations with larger extent).  Thus, stepping transitions between two-motor configurations are more likely to reduce the extent than to increase the extent. Therefore, over time, the two-motor ensembles would tend to have lower extent than a higher extent. 
This is what we observe in the \ref{fig:PerClusteredStates_2and3motor}(a), where with higher load force the two-motor configurations with smaller extent (i.e. the clustered states) appear to be more probable among the set of all two-motor configurations. An examination of the impact of load force on of the percentage share of clustered states for three-motor configurations, shown in Fig. \ref{fig:PerClusteredStates_2and3motor}(b), indicates similar trends.

\paragraph*{Impact}: With increasing load force on a multi-motor ensemble, relative configurations with small extents are preferred when multi-motor ensembles carry cargoes subjected to increasing load forces. The individual motors in such configurations have less variation in the magnitudes of their linkage extensions. Since the force on a single motor is directly proportional to the linkage extension, this leads to less variation in the load force exerted on the motors through the linkages. Thus, the load force on the cargo is shared more equally than in other configurations where the motors are farther apart and not clustered together. Since a lower force on a single motor leads to a higher probability of the motor taking a forward step (as long as the force is less than the stalling force $F_S$, see Appendix), ensembles in the prefered configurations have a higher probability of taking a forward step than in configurations with larger extent. An interesting possible consequence of such a configuration is that it allows for higher probability of cargo propagation.

As the load force is increased on the multi-motor ensembles, the preference for configurations with lesser extent does not always keep increasing with load force. This is evidenced by the trends seen in Fig. \ref{fig:PerClusteredStates_2and3motor}(a) and (b). Indeed, it is seen that above a certain value of load force, there is an abrupt departure from a preference to clustering of motors. For example, when a two-motor ensemble is subjected to a $F_{load} = 25 pN$ (Fig. \ref{fig:LoadForce_2motor}(c)), the most probable two-motor configuration is $M|||||M$, with a $46.5 \%$ share of the probability among all the $40$ two-motor configurations. When a three-motor ensemble is subjected to a $F_{load} = 50 pN$ (Fig. \ref{fig:LoadForce_3motor_High}), the most probable three-motor configuration is $MM-(11)-M$, with a $18.5 \%$ share of the probability among all the $1275$ three-motor configurations. An analysis of the load sharing in these configurations reveals non-intuitive insights. 

To understand how these specific prefered configurations emerge at high load forces, we first evaluate the forces exerted on each of the motors under the high load forces. In the two motor case with $F_{load}=25 pN$, the rearguard motor in the configuration $M|||||M$ handles $6.1 pN$ while the vanguard handles rest of the $18.9 pN$ load force. For three motor ensemble with $F_{load}=50 pN$, the rearguard motors in the configuration of $MM-(11)-M$ each handle $7.3 pN$ of the load force while the vanguard handles the rest $35.4 pN$ force. We notice a common attribute - the rearguard motor is subjected to a load force in the vicinity of its stalling force $F_S = 6 pN$ while the vanguard motor is loaded way beyond the stalling force. 

\subsection{Why is clustering abandoned at very high values of load force $F_{load}$? }

We again resort to using the basic properties of the single motor model to examine how such preferences emerge, as was utilized previously to explain the emergence of clustered states. Consider the example of two-motor ensemble subjected to a $F_{load} = 25 pN$ (Fig. \ref{fig:LoadForce_2motor}(c)). At any time instant during the journey of the cargo and ensemble towards its destination, the motor configuration would either be a one-motor configuration ($M$) or one of the $40$ two-motor configurations. Again, without loss of generality we can assume that the initial state of the ensemble is $M$. Here, the lone motor $M$ balances the $F_{load} = 25 pN$ which is significantly higher than its stalling force $F_S = 6pN$ and is thus stalled. From this initial condition, the only possibility is that the other floating motor reattaches to any location within $l_0$ distance of the cargo and leads to a two-motor configuration. Immediately after motor reattachment, since the linkage of the reattached motor is not stretched beyond $l_0$, the reattached motor bears no load force and becomes the rearguard motor. The vanguard motor still balances the $F_{load} = 25 pN$ and remains stalled while the rearguard motor is the only motor capable of taking a forward step. Once the ensemble is in such a two-motor configuration, the remaining possibilities are that either of the motors can detach or the rearguard motor can take a step forward. (1) If either motor detaches from the microtubule, the ensemble reverts back to the initial state $M$ after which there is a finite probability that it can return back to a two-motor configuration through reattachment. (2) If the rearguard motor takes a step, the extent of the configuration reduces. With every step of the rearguard motor, its linkage keeps extending further, thereby the force it balances also keeps increasing proportionally. Therefore, it begins to handle more and more of the load force, easing the burden on the vanguard motor which was initially balancing the entirety of the $F_{load} = 25 pN$. However, since the vanguard cannot step until the force it handles reduces from $25 pN$ to a value below the stalling force of $6 pN$, the stepping of the rearguard motor is the only stepping event possible. The rearguard motor keeps stepping forward until it cannot take a forward step i.e. the force on the rearguard motor equals or just exceeds the stalling force of $6 pN$ (the rearguard motor may not come to a halt in a configuration where it balances \textit{exactly} $6 pN$ for load force, due to the discretization of the microtubule in dimers of finite $8 nm$ length). Therefore, the most likely two-motor configuration for a two-motor ensemble under $F_{load} = 25 pN$ is $M|||||M$ (Fig. \ref{fig:LoadForce_2motor}(c)), where the rearguard motor is at or just beyond stalling condition and the vanguard motor handles the rest of the $F_{load}$. Similar explanation holds for the three-motor ensemble subjected to a $F_{load} = 50 pN$ (Fig. \ref{fig:LoadForce_3motor_High}), with the most probable three-motor configuration being $MM-(11)-M$. This also explains the abrupt drop in the percentage share of clustered states in Fig. \ref{fig:PerClusteredStates_2and3motor}(a) and (b) above load forces of a certain value. At these high values of load forces, the most likely configuration has an extent more than two microtubule locations (which is how we have defined \textit{clustered states}) and the vanguard and rearguard motors are stalled. Thus, the rearguard motor is incapable of taking a step and reducing the extent any further, making the clustered states improbable and leading to an abrupt drop in their percentage share.


\paragraph*{Impact}: At high load forces the preferred configuration of motors in a multi-motor ensemble has a single motor in the vanguard position with all the rest of the motors being rearguard motors. The extent of the prefered configuration is such that the rearguard motors are loaded at or just beyond their stall force $F_s$ and the vanguard motor bears the remaining load, whose value is more than $F_s$. Such an orientation can possibly offer unique advantages while dealing with high load forces. Circumstances of cargo being subjected to high load force relate to infrequent yet possibly debilitating phenomenon inside the cells; such as an unanticipated obstacle along the cargo travel path or an encounter with an oppositely directed cargo. Occurrences of such nature are most likely sudden events that do not lead to a progressive loading of the load force on the cargo, but an unexpected spike in $F_{load}$. Based on our analysis, the preferred orientation is such that as few motors as possible (i.e. a single vanguard motor) are loaded beyond the stall force. In such an arrangement, once the transient loading event goes away and $F_{load}$ starts to subside, it would take little reduction in $F_{load}$ to reduce the force on each of the rearguard motors to fall below $F_S$. Thereby, this enabling the rearguard motors to take a forward step and propel the cargo forward (since force on each motor now falls below $F_s$). The prefered configuration offers the maximum number of motors that can be mobilized by the least reduction in $F_{load}$, while keeping the load force on the one vanguard motor as low as possible. In contrast, for configurations where the rearguard and vanguard motors are closer (like a clustered configuration preferred for lower $F_{load}$), the non-vanguard motors would be taking a load higher than $F_s$ and there would be less number of motors loaded \textit{just above} their stall force. Thus a larger reduction of load force would be necessary to enable the same numbers of motors to be able to step forward, as in the previous configuration. Another possible advantage is that, for the same reduction in $F_{load}$, the preferred orientation enables the maximum number of motors (i.e. all but one vanguard) to be able to walk while maintaining the load force on the vanguard motor as low as possible. Thus, this orientation ensures that the least amount of reduction in the high load force is needed to enable the resumption of motor motion and forward propagation of cargo towards the destination.

\section{Conclusions}

Using a semi-analytic Markov model to analyze intracellular cargo transport by teams of finite motors, we prove that the motors orient themselves according to \textit{unique steady state distributions} irrespective of the initial orientation. It demonstrates the robustness of the intracellular transport mechanism. Analyzing how the distribution is impacted by external factors reveals interesting coordination mechanisms. For a team of multiple Kinesin-I motors at reduced ATP concentrations, higher number of motors prefer to remain attached to the microtubule during the course of cargo transport. This contributes to enhanced ensemble run-length despite despite reduced  velocitiy. Furthermore, an increase in hindering load force on the cargo is tackled by the motors \textit{'clustering'} together. However, at very high load forces the motors abandon clustering and adopt configurations that prefer anchoring the cargo and immediate cargo translocation once the large loading event has subsided. Our approach provides unique insights into the team behaviors of molecular motors, with the analytical approach suitable to study transport mechanisms adapted by teams of other motors, such as dynein, myosin or even heterologous ensembles.  

\section{Appendix}
\subsection*{Probability Rates for Kinesin}

The results obtained in this article and the supplementary material correspond to an instantiation of the semi-analytic methodology for the motor protein Kinesin-I. The external inputs for the method are a triplet of probabilities - probability of stepping, detachment and reattachment. The probability rates are determined based on existing studies \cite{klumpp2015molecular,kunwar2010robust,deville2008regularity,svoboda1994force}. 

\subsection*{Notations}

\begin{itemize}
	\item $X_C$ - mean position of the cargo
	\item $k_{on}$ - Rate constant for the binding of an ATP molecule by the Kinesin head
	\item $k_{off}$ - Rate constant for the unbinding of an ATP molecule from the Kinesin head
	\item $ATP$ - Adenosine Triphosphate
	\item $ADP$ - Adenosine Diphosphate
	\item $\mathcal{M}$ - Motor
	\item $k_{cat}$ - Rate constant for ATP hydrolysis
	\item $k_m$ - Michaelis-Menten constant ()given by $k_m =\frac{k_{cat}+k_{off}}{k_{on}}$)
	\item $P_S$ - Probability of motor stepping
	\item $\eta$ - Efficiency with which a free motor head binds to a location on the microtubule
	\item $F$ - Force exerted on the motor by the cargo
	\item $k_{0,off}$ - Backwards reaction rate of hydrolysis under no force condition i.e. $F = 0$
	\item $K_b$ - Boltzmann's constant
	\item $T$ - Absolute temperature
	\item $F_S$ - Stalling force of the motor
	\item $\mathcal{L}$ - Processivity of the motor i.e. the average distance traveled by the motor before it's detachment from the microtubule
	\item $d$ - Average stepping length of a motor
	\item $P_D$ - Probability of detachment of the motor from the microtubule
	\item $d_l, \delta_l, A, B$ - Experimentally determined parameters from \cite{kunwar2008stepping, schnitzer2000force}. 
	\item $\sigma_{th}$ - variance of the cargo position at steady state, used to incorporate the effect of thermal fluctuations on the cargo position.
	
	
\end{itemize}

\subsection{Probability of Stepping, per second} 

The Kinesin-I motor protein takes a forward step on the microtubule by hydrolyzing a molecule of ATP \cite{schnitzer1997kinesin} in the following manner:
\begin{equation*}
\mathcal{M}+ATP \overset{k_{on}}{\underset{k_{off}}{\rightleftharpoons}} \mathcal{M}~ATP \xrightarrow{k_{cat}} \mathcal{M}+ADP+p_{i}+\Delta e,
\end{equation*}
where $\Delta e$ is the energy released from the ATP hydrolysis reaction. In \cite{meyhofer1995force} the probability rate of stepping $P_S$ is obtained based on the ATP hydrolysis rate predicted using the Michaelis-Menten dynamics. Here,  a free motor head binds to a location on the microtubule with an efficiency $\eta$. $P_S$ is expressed as,
\begin{equation}
P_S=\frac{k_{cat}[ATP]}{[ATP]+k_m}\eta,
\end{equation}
where $k_m =\frac{k_{cat}+k_{off}}{k_{on}}.$

In \cite{kunwar2008stepping}, the force $F$ exerted on the motor by the cargo is assumed to impact the motor dynamics by altering the efficiency $\eta$ as,

\begin{equation}\label{eq:efficiency_calc}
\eta(F) = \left\{ 
\begin{array}{l l}
1 & \quad \text{if}~F=0,\\
1-{(\frac{F}{F_s})}^2 & \quad \text{if}~0<F<F_s,\\
0 & \quad \text{otherwise}.\\
\end{array} \right.
\end{equation}

In \cite{kunwar2008stepping}, it is further assumed that the kinetics of the ATP hydrolysis are also impacted by the force $F$ by affecting $k_{off}$ as,
\begin{equation*}
k_{off}(F)=k_{0,off}e^{\frac{F d_l}{K_b T}},
\end{equation*}
Here, $k_{0,off}$ is the backward reaction rate of hydrolysis when $F = 0$, $T$ is the absolute temperature, $K_b$ is the Boltzmann constant and $\delta_l$ is a parameter determined experimentally. Incorporating the impact of all the above variables, the probability rate of stepping for Kinesin-I motor under a constant force $F$ is given by,
\begin{equation}\label{Pstep_H}
P_S(F)=\frac{k_{cat}[ATP]}{[ATP]+{\frac{k_{cat}+k_{off}(F)}{k_{on}}}}\eta(F).
\end{equation}

If we assume that the cargo position follows a truncated Gaussian distribution with the probability density, for $|X|<3\sigma_{th}$,
\begin{equation}\label{Eq_position_pdf}
\Phi(X)=\frac{e^{-\frac{X^2}{2\sigma_{th}^2}}}{2\int_{0}^{3\sigma_{th}} e^{-\frac{X^2}{2\sigma_{th}^2}} dX},
\end{equation}
the transition rate of stepping can be determined by averaging over the position of the cargo as
\begin{equation}\label{transRate_step}
\begin{aligned}
\lambda(Z^{step}, Z) = z_i \int_{X_C(Z) - 3\sigma_{th}}^{X_C(Z) + 3\sigma_{th}} & P_S(F(X-a_i)) \\ & \Phi(X-X_C(Z)) dX
\end{aligned}
\end{equation}
where $Z^{step}$ is the absolute configuration obtained after $z_i$ motors at the $i^{th}$ location in $Z$ take a forward step and $a_i$ is the position of the $i^{th}$ location. Here, the transition rate is proportional to the numbers of motors at the location.

\subsection{Probability of Detachment, per second}

Kinesin motors are known to take a certain number of steps on the microtubule during their ATP fuled movement before disengaging from the microtubule. The The \textit{processivity} ($\mathcal{L}$) denotes the average distance moved by the motor before it's detachment from the microtubule. Using \cite{schnitzer2000force}, the processivity $\mathcal{L}$ is given by,
\begin{equation*}
\mathcal{L}=\frac{d [ATP]Ae^{-F\delta_l / K_bT}}{[ATP]+B(1+Ae^{-F\delta_l / K_bT})},
\end{equation*}
Here, the parameters $A$,$B$ and $\delta_l$ are  determined experimentally. Since processivity denotes the average distance traversed by the motor before it's detachment from the microtubule, we can provide a relation between the probability rate of stepping $P_S$ and detachment $P_D$ as
\begin{equation}\label{Pdetach}
\frac{P_S(F)}{P_{D}(F)}=\frac{\mathcal{L}}{d}.
\end{equation}

So long as the motor is not stalled ( i.e. $F<F_s$), the probability rate of detachment $P_{D}(F)$ is given by,
\begin{equation}\label{Pdetach_W}
P_{D}(F)=\frac{[ATP]+B(1+Ae^{-F\delta_l / K_bT})}{[ATP]Ae^{-F\delta_l / K_bT}} P_S(F).
\end{equation}

If the force on the motor equals  or exceeds its stalling force (i.e. $F\geq F_s$), a constant detachment rate is assumedbased on \cite{kunwar2008stepping} as,
\begin{equation*}
P_{D}(F)=P_{back}=2/sec.
\end{equation*}

Similar to the description in equation (\ref{transRate_step}) the transition rate for detachment event can be described as,
\begin{equation}\label{transRate_det}
\begin{aligned}
\lambda(Z^{det}, Z) = z_i \int_{X_C(Z) - 3\sigma_{th}}^{X_C(Z) + 3\sigma_{th}} & P_D(F(X-a_i)) \\ & \Phi(X-X_C(Z)) dX
\end{aligned}
\end{equation}
where $Z^{det}$ is the absolute configuration obtained after $z_i$ motors at the $i^{th}$ location in $Z$ detach from the microtubule and $a_i$ is the position of the $i^{th}$ location. 


\subsection{Probability of Attachment, per second}

It was found experimentally in \cite{beeg2008transport,leduc2004cooperative} that the probability of attachment of the kinesin motor to the microtubule is $P_A \approx 5/sec$. In the model, we assume that a motor linked to the cargo can attach to all the locations on the microtubule that are accessible without stretching it's linkage. All of these locations are assumed to be equally likely. 

\paragraph*{Numerical parameters for Kinesin: } 
The numerical parameters considered in this article for the Kinesin-I motor are similar to the ones used in \cite{kunwar2008stepping}, which are themselves adopted from experimentally and emperically backed studies. Specifically $k_{on}=2.10^6 ~M^{-1}s^{-1}$, $k_{off}=55~ s^{-1}$, $k_{cat}=105~s^{-1}$, $F_s=6~pN$, $d=8~nm$, $d_l=1.6~nm$, $\delta_l=1.3~nm$, $A=107$, $B=0.029 ~\mu M$, $[ATP]=2~mM$, $T=300K$ and $K_{el}=0.32~pN/nm$. 

\bibliographystyle{apsrev4-1}

\bibliography{biblio}

\end{document}